# Low-noise, 2-W average power, 112-fs Kerr-lens mode-locked Ho:CALGO laser at 2.1 µm

WEICHAO YAO,[1,*] YICHENG WANG,[1] SHAHWAR AHMED,[1] MARTIN HOFFMANN,[1] MARCEL VAN DELDEN,[2] THOMAS MUSCH,[2] AND CLARA J. SARACENO[1]

[1]*Photonics and Ultrafast Laser Science, Ruhr Universität Bochum, Universitätsstraße 150, 44801 Bochum, Germany*
[2]*Institute of Electronic Circuits, Ruhr Universität Bochum, Universitätsstraße 150, 44801 Bochum, Germany*
*Corresponding author: weichao.yao@ruhr-uni-bochum.de

**We report on an in-band pumped soft-aperture Kerr-lens mode-locked Ho:CALGO bulk laser at 2.1 µm, generating 2 W of average power with 112 fs pulses at 91-MHz repetition rate. To the best of our knowledge, this is the highest average power from a 100-fs class mode-locked laser based on a $Tm^{3+}$ or $Ho^{3+}$ doped bulk material. We show that the laser has excellent noise properties with an integrated relative intensity noise of 0.02% and a timing jitter of 950 fs (RMS phase noise 0.543 mrad) in the integration interval from 10 Hz to 10 MHz. The demonstrated combination of high average power, short pulses, and low-noise make this an outstanding laser source for spectroscopy and many other applications at 2.1 µm.**

Ultrafast high-power 2 µm lasers are currently the topic of intense investigation in the laser community, mostly motivated by the advantages of longer driving wavelengths in the fields of material processing [1], spectroscopy [2], and nonlinear conversion. Among the different classes of laser systems that are currently being developed in this wavelength region, bulk solid-state lasers directly emitting at this wavelength are the most attractive solutions due to their simplicity and potential for high efficiency. Several types of laser systems are currently being explored: $Tm^{3+}$, $Ho^{3+}$, or $Tm^{3+}/Ho^{3+}$ (co-)doped bulk materials and Cr:ZnS(e) which emits at 2.4 µm. Mode-locked lasers based on $Tm^{3+}$ or $Tm^{3+}/Ho^{3+}$ co-doped bulk materials [3-6] have mostly been explored for their potential to generate short pulses less than 50 fs albeit at low average power [3, 4]. On the other hand, watt-level average powers have been achieved from these lasers but with longer pulse duration of several hundreds of femtoseconds [5, 6]. Using $Ho^{3+}$ doped materials, the pulse duration remains at the picosecond level in bulk lasers, with femtosecond operation only demonstrated for thin-disk and fiber laser gemoetries [8, 9]. Concerning Cr:ZnS(e) mode-locked lasers, 2 W of average power with 67 fs of pulse duration directly from the oscillator has been reported [7]. Further scaling is possible here using MOPA architectures, however further scaling of the oscillators themselves beyond the few-watt level appears difficult due to the thermal properties of the crystals [10]. In addition, we note the central wavelength of Cr:ZnS(e) is much longer than all of the above mentioned lasers, leading to different application possibilities. In this respect, Holmium lasers, typically emitting around 2.1 µm have well-known advantages in this spectral region because they fall in a "high transmission window" in this spectral region.

Generally, it is challenging to combine high average power and short pulses in mode-locked lasers. This is typically due to conflicting properties of the laser gain material in terms of thermal conductivity and gain bandwidth; pump power and pump beam quality available, but also mode-locking and corresponding loss mechanisms. In this regard, mode-locked lasers based on $Ho^{3+}$-doped materials can still offer a promising route because of typically excellent thermal properties and simple quasi-three-level energy scheme of the material, enabling to generate high average powers with a single-mode Tm fiber laser pump source. The problem in past studies is the narrow gain from most commonly used hosts, thus pulse durations remained long. Recently, we showed that $Ho^{3+}$-doped $CaGdAlO_4$ (Ho:CALGO) is a competitive candidate material for circumventing this issue and achieving short pulses and high-power mode-locking. Using SESAM mode-locking we could achieve up to 8.7 W of average power with 369 fs pulses, which is the highest power so far achieved with bulk mode-locked lasers at 2.1 µm - confirming this potential [11]. The pulse duration in this experiment was limited by the low modulation depth of the saturable absorber. However, the broadband gain spectrum of Ho:CALGO (>50 nm in $\sigma$-polarization, inversion ratio 0.32) [12] induced by a disordered structure should support 100-fs level mode-locked pulses. This could be accessed with faster and more broadband saturable absorbers, such as Kerr-lensing.

In this work, we demonstrate the first Kerr-lens mode-locked Ho:CALGO laser, delivering 2 W of average power with 112 fs pulse duration, which is to the best of our knowledge the highest power 100-fs-class mode-locked bulk laser from a $Tm^{3+}$- or $Ho^{3+}$-doped material, and the shortest pulses achieved with a $Ho^{3+}$-based material. Additionally, motivated by the potential of this laser system for future spectroscopy applications, we measured the noise properties (relative intensity noise (RIN) and phase noise) at 2-W average power, showing that this laser exhibits exceptional noise and excellent long-term stability.

The experimental setup of the KLM Ho:CALGO laser is shown in Fig. 1. A 10-mm long, ***a***-cut 3.1 at.% doped Ho:CALGO crystal was used as a gain medium. Its clear aperture was 4 mm × 4 mm, and both end surfaces of the crystal were anti-reflection-coated for the wavelength range from 1900 nm to 2200 nm. The crystal was water-cooled at 16 °C. We used an asymmetrical resonator, in which the two concave mirrors M1 and M2 have radii of curvature ($R_{OC}$) of -200 mm and -300 mm, respectively. The continuous-wave (CW) mode radius in the crystal was calculated to be ~110 µm. For pumping, we used a single-mode, unpolarized 1940-nm Tm fiber laser. To introduce a soft aperture effect, the beam was focused with a slightly smaller cavity mode radius of ~105 µm in the crystal, by means of the in-coupling mirror IM1. All cavity mirrors are highly reflective for both laser and pump, except for the mirrors IM1 and IM2, which exhibit high transmission for the pump and high reflectivity for the laser. Hence, a second pump pass through the crystal is intentionally avoided by coupling the pump out through IM2. In our experiment, the maximum incident pump power was set to 15.7 W to reduce the risk of crystal damage, due to the lower conversion efficiency of KLM compared with that of SESAM mode-locking, and the smaller pump/cavity modes in the crystal [11]. During laser operation, the absorbed pump power was determined by the difference between incident pump power and leaked pump power from IM2. To optimize the output power, five different output couplers (OCs) with output transmissions of 1%, 2%, 3%, 4%, and 5% were used. Concerning dispersion management in the cavity, the 10 mm long CALGO crystal provides a total round-trip group delay dispersion (GDD) of ~ 1100 fs$^2$ at 2.15 µm for $\sigma$-polarization [13], and we add an additional ~ 1000 fs$^2$ of round-trip GDD using a dispersive mirror (DM) to optimize and stabilize mode-locking. The resulting repetition rate in this configuration amounts to 91 MHz.

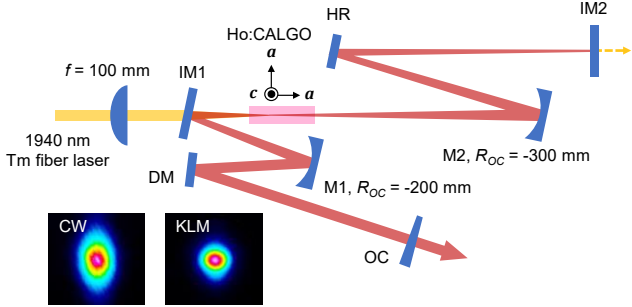

**Fig. 1.** Experimental setup of the KLM Ho:CALGO laser. IM1-2: input mirrors; HR: high-reflectivity coated mirror; DM: dispersive mirror, -500 fs$^2$ per bounce; OC: output coupler. The arrows next to the crystal are the crystal axes directions. Inset: beam profile of continuous-wave (CW) and Kerr-lens mode-locking (KLM) at ~1.2 m away from OC ($T_{OC}$ = 5%) at the maximum incident pump power.

We first explored CW operation to optimize the cavity with 1% $T_{OC}$, since the laser threshold is lower. A maximum output power of 1.5 W was achieved at the incident pump power of 6.8 W (single-pass absorption: 68%), corresponding to an optical-to-optical efficiency of 22.1% with respect to the incident pump power. The laser wavelength was 2135 nm in the $\sigma$-polarization direction. In addition, we found that the polarization direction does not change with any of the other four OCs in the following experiments, and also not in the mode-locking experiments.

To achieve mode-locking, the cavity was adjusted towards the edge of the stability zone: the concave mirror M2 was moved towards the crystal by ~1.4 mm. In this case, the CW output power dropped to 0.59 W, and the wavelength shifted to 2127 nm because of a high inversion ratio driven by the increased cavity loss. Mode-locking was started by slightly pushing the end mirror IM2. This leads to an output power increase to 0.65 W, corresponding to an optical-to-optical efficiency of 9.6% with respect to the incident pump power (single-pass absorption: 63%). The laser remains in stable mode-locking operation for 6 W to 6.8 W of incident pump power. At higher pumping power, CW breakthrough will appear in the laser spectrum indicating the beginning of mode-locking instabilities. Figure 2 shows the output performance of the KLM Ho:CALGO laser with 1% $T_{OC}$ at the highest output power. The fitted spectral bandwidth (full width at half maximum, FWHM, $\Delta\lambda$) was 47.1 nm at 2151.3 nm, as shown in Fig. 2(a). The autocorrelation trace is illustrated in Fig. 2(b). Assuming a sech$^2$-shape pulse profile, the fitted pulse duration (FWHM, $\Delta\tau$) was 104 fs. The corresponding time-bandwidth product (TBP) is 0.317, which is very close to the Fourier-transform-limited value of 0.315. Compared with the laser spectrum using SESAM mode-locking [11], the current laser spectrum is strongly broadened and shifted to a longer wavelength, which can be attributed to the stronger self-phase modulation in the crystal (intra-cavity peak intensity ~17 GW/cm$^2$) and effective gain reduction with a shorter pulse duration, enabling the generation of the shortest pulse duration to date from a mode-locked Ho-based laser system.

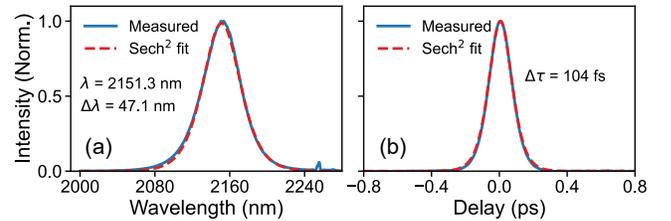

**Fig. 2.** Output performance of the KLM Ho:CALGO laser with 1% $T_{OC}$ at 0.65-W average output power. (a) Laser spectrum. (b) Autocorrelation trace.

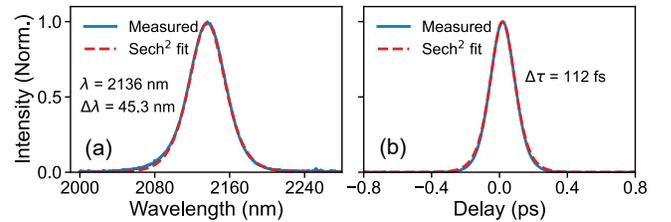

**Fig. 3.** Output performance of the KLM Ho:CALGO laser with 5% $T_{OC}$ at 2-W average output power. (a) Laser spectrum. (b) Autocorrelation trace.

By increasing the transmission of the output coupler, the average output power of the laser is increased with a nearly unchanged pulse duration. To achieve this, the cavity was again adjusted toward the edge of the stability region in order to slightly increase the modulation depth and to ensure the shortest pulse duration. Hence, we were able to increase $T_{OC}$ up to 5% with stable mode-locking. The results with different output coupler transmissions are shown in Table 1. With 2%, 3%, and 4% $T_{OC}$s, the average power amounted to 1.2 W, 1.5 W, and 1.75 W, and pulse durations were 104 fs, 106 fs, and 109 fs, respectively. With 5% $T_{OC}$, the average output power was scaled to 2 W at an incident pump power of 15.7 W (single-pass absorption: 51.5%), corresponding to an optical-to-optical efficiency of 12.7%. The output performance of the KLM Ho:CALGO laser with 5% $T_{OC}$ is shown in Fig. 3. The center wavelength shifts from 2122 nm in CW to 2136 nm in KLM, with a fitted spectral bandwidth of 45.3 nm (FWHM). The fitted pulse duration from the autocorrelation trace is 112 fs (FWHM), with a TBP of 0.333, indicating slightly chirped pulses. Higher output power or shorter pulse duration were limited by the onset of CW breakthrough in the laser spectrum. To the best of our knowledge, this is the highest average power so far from a 100-fs scale mode-locked Tm or Ho bulk laser.

**Table 1. KLM Ho:CALGO laser results with different $T_{OC}$s. $P_{in}$: incident pump power. $P_{out}$: average output power. $\eta_{abs}$: single-pass absorption. $\eta_{oto}$: optical-to-optical efficiency. TBP: time-bandwidth product.**

| $T_{OC}$ (%) | 1 | 2 | 3 | 4 | 5 |
|---|---|---|---|---|---|
| $P_{in}$ (W) | 6.8 | 9.8 | 10.5 | 12 | 15.7 |
| $P_{out}$ (W) | 0.65 | 1.2 | 1.5 | 1.75 | 2 |
| $\eta_{abs}$ (%) | 63 | 56.7 | 55.3 | 55.2 | 51.5 |
| $\eta_{oto}$ (%) | 9.6 | 12.2 | 14.3 | 14.6 | 12.7 |
| $\lambda$ (nm) | 2151.3 | 2145.2 | 2143.6 | 2140 | 2136 |
| $\Delta\lambda$ (nm) | 47.1 | 46.9 | 46.5 | 45.5 | 45.3 |
| $\Delta\tau$ (fs) | 104 | 104 | 106 | 109 | 112 |
| TBP | 0.317 | 0.317 | 0.322 | 0.325 | 0.333 |

By increasing the $T_{OC}$ further to 7%, we observed CW-emission of shorter wavelengths at 2080 nm ($\pi$-polarization) in the laser spectrum, which was difficult to suppress completely because of a high cavity loss, seemingly limiting further power scaling by increasing the output transmission. However, continuing to enlarge the mode radius in the crystal with 5% $T_{OC}$ will be beneficial to increase the output power while keeping the pulse duration [14], but with a higher mode-locking threshold. In this case, the possible risk of crystal damage can be mitigated by optimizing the crystal size and heat sink in the future. At 2 W of average output power, the output peak power is 173 kW, while the intra-cavity peak intensity in the crystal is ~10 GW/cm², which inevitably leads to a significant reduction of the laser mode size, after KLM is established. The beam profiles at ~1.2 m away from the OC were measured with a micro-bolometer camera at the maximum incident pump power, corresponding to a CW power of 1 W and a mode-locked average power of 2 W. The measured diameter reduced from 2.23 mm × 3.59 mm (CW) to 2.03 mm × 1.95 mm (KLM), as shown in Fig. 1.

The characterization of the mode-locked pulses is shown in Fig. 4. Because of the nearly identical measurements for the different $T_{OC}$s, only the results at the maximum output power, i.e., 2 W, are presented here however same mode-locking stability was observed at the other data points as well. At the maximum output power, we scanned the autocorrelation trace of the pulse with a 16-ps scale, as shown in Fig. 4(a), and the pulses were measured with a 12.5-GHz fast photodiode and recorded with a 25-GHz sampling oscilloscope (PicoScope 9000, Pico Tech.), see Fig. 4(b). There is a weak signal at ~6 ns, which should be caused by electromagnetic interference. The time delay between two pulses is ~11 ns corresponding to the round-trip time of the cavity. These measurements show no indication of harmonic mode-locking or multi-pulsing. Moreover, Fig. 4(c) and Fig. 4(d) show the radio frequency spectra in a span of 1 GHz and the fundamental beat note, respectively, each measured with a 12.5-GHz fast photodiode and recorded by a radio frequency spectrum analyzer. The harmonic beat notes exhibit nearly the same intensity further showing stable mode-locking without modulation. The fundamental repetition rate beat note has a signal-to-noise ratio of 60 dBc, and indicates no Q-switching instabilities.

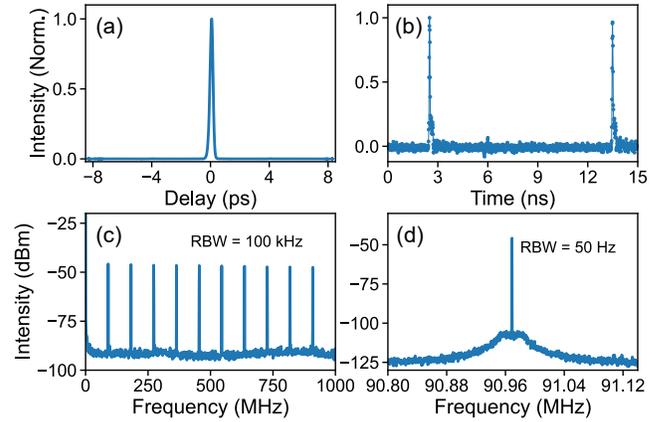

**Fig. 4.** Characterization of the mode-locked laser pulse train. (a) 16-ps scale autocorrelation scan. (b) Sampling oscilloscope measurement. The weak signal at ~6 ns is a spurious signal from electromagnetic interference. (c) Radio frequency spectrum in a span of 1 GHz. RBW: resolution bandwidth. (d) Radio frequency spectrum of the fundamental beat note.

We also characterized the stability of our mode-locked laser by performing amplitude and phase noise measurements. We note that all the measurements above were measured with the laser running in a normal air enclosure, without optimizing the mechanical design for stability. The laser was operated at 2 W of average output power, and the pulses were detected with our 12.5-GHz fast-photodiode and analyzed with a 50-GHz phase noise analyzer (Rohde & Schwarz FSWP50). For the amplitude noise measurement, the measurement was performed on the 91-MHz fundamental repetition rate beat note. As a reference, the amplitude noise of our continuous-wave Tm-fiber laser was measured at baseband as well at the corresponding pump power, i.e., 15.7 W. The results are shown in Fig. 5. For the Tm-fiber laser, besides the relaxation-oscillation induced broad peak at ~350 kHz and longitudinal mode-beating induced peak signal at ~8.3 MHz, the laser's power spectral density (PSD) has no intensity noise peaks, leading to an integrated RIN of 0.65% in the integration interval from 10 Hz to 10 MHz. For the KLM Ho:CALGO laser, the amplitude noise generated at low frequency (<1 kHz) is dominated by technical noise, which can be easily reduced by mechanical improvements and slow feedback on one of the cavity mirrors. The noise peak at ~22 kHz can be attributed to relaxation-oscillations; in fact the offset frequency changes for different pump powers. At high frequencies (>22 kHz), the RIN PSD is as expected significantly attenuated by the lifetime of the gain medium and is close to the background noise floor. The integrated RIN of the KLM Ho:CALGO laser in the integration interval from 10 Hz to 10 MHz is 0.02%, which is even lower than that of low-noise Cr:ZnSe and Cr:ZnS lasers [15-17] operating at much lower power (i.e. Int. RIN of 0.05% in [10 Hz, 5 MHz] for Cr:ZnS laser at 0.55 W of output power [15]). This outstanding low-noise performance of the laser is attributed to the low-noise properties of the single-mode pump laser at low frequencies. In fact, most of the contributions to the RIN of the pump are located at higher than 100 kHz frequencies, which are filtered by the gain lifetime. Furthermore, the long-term average output power at 2 W was also measured with a power meter continuously for 1 hour giving an RMS stability of 0.06% indicating also excellent long-term stability.

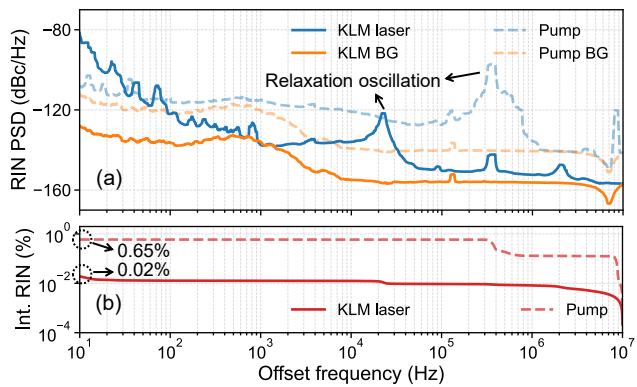

**Fig. 5.** Amplitude noise of the Ho:CALGO laser system at 2-W average output power. (a) Power spectral density for relative intensity noise. (b) Integrated RIN. BG: background noise floor.

The phase noise measurement was performed on the 10th harmonic to improve the measurement sensitivity by around 18 dB. Figure 6 shows the phase noise PSD and integrated timing jitter of the KLM Ho:CALGO laser at 2 W of average output power. Similar to the behavior of amplitude noise in Fig. 5, the phase noise PSD increases only at low frequencies (<1 kHz) and at the frequency of relaxation oscillation (~22 kHz). The integrated timing jitter increases to 30 fs (RMS phase noise 0.017 mrad) at 15 kHz because of relaxation-oscillation. This value increases to 36 fs (RMS phase noise 0.02 mrad) in the range from 0.7 kHz to 10 MHz before mechanical noise dominates the noise properties. The integration over the entire range from 10 Hz to 10 MHz leads to a timing jitter of 950 fs (RMS phase noise 0.543 mrad). These results demonstrate that our KLM Ho:CALGO laser exhibits much lower phase noise compared with mode-locked Cr:ZnS and Cr:ZnSe lasers [15-17], thus is a very promising candidate for further jitter stabilization, and even in the future for super-continuum generation and full comb stabilization.

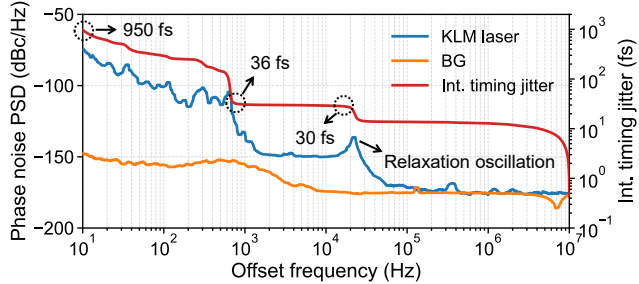

**Fig. 6.** Phase noise PSD and integrated timing jitter of the KLM Ho:CALGO laser at 2-W average output power.

In conclusion, we have successfully demonstrated a mode-locked Ho:CALGO laser with high average output power and ultrashort pulses employing the soft-aperture KLM mechanism. In optimized conditions, we obtained an average power of 2 W, with 112-fs pulse duration, which represents the highest power so far obtained with 100-fs pulses in this spectral range; and the shortest pulses obtained with Ho. At the highest power level, the laser shows very low noise and timing jitter in the absence of active stabilization, which confirms the large potential of Ho:CALGO for high-power, short pulse 2.1 μm lasers, and indicates this is a promising technology for spectroscopy and other applications requiring high-power low noise ultrafast lasers.

**Funding.** European Research Council (805202); Deutsche Forschungsgemeinschaft (390677874, 287022738 TRR 196).

**Acknowledgments.** This project was funded by the Deutsche Forschungsgemeinschaft (DFG) under Germany's Excellence Strategy - EXC 2033 - 390677874 – RESOLV and also under Project-ID 287022738 TRR 196 (SFB/TRR MARIE). These results are part of a project that has received funding from the European Research Council (ERC) under the European Union's Horizon 2020 research and innovation programme (grant agreement No. 805202 - Project Teraqua). We acknowledge support by the MERCUR Kooperation project "Towards an UA Ruhr ultrafast laser science center: tailored fs-XUV beam line for photoemission spectroscopy." W. Yao acknowledges financial support from the Alexander von Humboldt Foundation through a Humboldt Research Fellowship.

**Disclosures.** The authors declare no conflicts of interest.

**Data availability.** Data underlying the results presented in this paper are not publicly available at this time but may be obtained from the authors upon reasonable request.

### References

1. R. A. Richter, N. Tolstik, S. Rigaud, P. D. Valle, A. Erbe, P. Ebbinghaus, I. Astrauskas, V. Kalashnikov, E. Sorokin, and I. T. Sorokina, J. Opt. Soc. Am. B **37**(9), 2543-2556 (2020).
2. W. Song, D. Okazaki, I. Morichika, and S. Ashihara, Opt. Express **30**(21), 38674-38683 (2022).
3. A. Suzuki, C. Kränkel, and M. Tokurakawa, Opt. Express **29**(13), 19465-19471 (2021).
4. N. Zhang, Q. Song, J. Zhou, J. Liu, S. Liu, H. Zhang, X. Xu, Y. Xue, J. Xu, W. Chen, Y. Zhao, U. Griebner, and V. Petrov, Opt. Lett. **48**(2), 510-513 (2023).
5. M. Tokurakawa, E. Fujita, and C. Kränkel, Opt. Lett. 42(16), 3185-3188 (2017).
6. N. Zhang, Z. Wang, S. Liu, W. Jing, H. Huang, Z. Huang, K. Tian, Z. Yang, Y. Zhao, U. Griebner, V. Petrov, and W. Chen, Opt. Express **30**(13), 23978-23985 (2022).
7. S. B. Mirov, V. V. Fedorov, D. Martyshkin, I. S. Moskalev, M. Mirov, and S. Vasilyev, IEEE J. Sel. Top. Quantum Electron. **21**(1), 292–310 (2015).
8. J. Zhang, K. F. Mak, and O. Pronin, IEEE J. Sel. Top. Quantum Electron. **24**(5), 1102111(2018).
9. P. Li, A. Ruehl, U. Grosse-Wortmann, and I. Hartl, Opt. Lett. 39(24), 6859-6862 (2014).
10. I. Moskalev, S. Mirov, M. Mirov, S. Vasilyev, V. Smolski, A. Zakrevskiy, and V. Gapontsev, Opt. Express **24**(18), 21090–21104 (2016).
11. W. Yao, Y. Wang, S. Tomilov, M. Hoffmann, S. Ahmed, C. Liebald, D. Rytz, M. Peltz, V. Wesemann, and C. J. Saraceno, Opt. Express **30**(23), 41075-41083 (2022).
12. Y. Wang, P. Loiko, Y. Zhao, Z. Pan, W. Chen, M. Mero, X. Xu, J. Xu, X. Mateos, A. Major, M. Guina, V. Petrov, and U. Griebner, Opt. Express **30**(5), 7883–7893 (2022).
13. P. Loiko, P. Becker, L. Bohatý, C. Liebald, M. Peltz, S. Vernay, D. Rytz, J. M. Serres, X. Mateos, Y. Wang, X. Xu, J. Xu, A. Major, A. Baranov, U. Griebner, and V. Petrov, Opt. Lett. **42**(12), 2275-2278 (2017).
14. J. Brons, V. Pervak, D. Bauer, D. Sutter, O. Pronin, and F. Krausz, Opt. Lett. **41**(15), 3567-3570 (2016).
15. J. Heidrich, A. Barh, S. L. Camenzind, B. Willenberg, M. Gaulke, M. Golling, C. R. Phillips, and U. Keller, IEEE J. Quantum Electron., **59**(1), 1300107(2023).
16. A. Barh, B. Ö. Alaydin, J. Heidrich, M. Gaulke, M. Golling, C. R. Phillips, and U. Keller, Opt. Express **30**(4), 5019-5025 (2022).
17. Y. Wang, T. T. Fernandez, N. Coluccelli, A. Gambetta, P. Laporta, and G. Galzerano, Opt. Express **25**(21), 25193-25200 (2017).